\documentclass[prd,aps]{revtex4-2}

\usepackage{graphicx}

\usepackage{color}

\begin{document}

\title{Superposition model test in small CORSIKA shower simulations}

\author{Tadeusz Wibig}
\affiliation{Department of Theoretical Physics, Faculty of Physics and Applied Informatics, University of Lodz,  \\ Pomorska 149/153, PL-90-236 \L \'od\'z, Poland.}
\email{tadeusz.wibig@fis.uni.lodz.pl}

\vspace{10pt}

\date{\today}
\begin{abstract}
The calculation of the flux of particles reaching the earth's surface, as well as cases of simultaneous arrival at the observation level of groups of particles -- very small Extensive Air Showers, require simulation calculations and taking into account a much greater extent of fluctuations of the development of the showers generated by very low energy cosmic ray particles, much greater than usual for showers with energies near the ``knee'' and above due to the very steep cosmic ray energy spectrum.
These fluctuation depend significantly on the mass number of the primary cosmic ray particle. Nuclei are sets of nucleons and it seems an obvious approximation to assume that the extensive air showers initiated in the upper atmosphere by atomic nuclei are a convolution of the showers generated by single nucleons. This is called the superposition hypothesis. In this paper, we test this superposition model. We will show, using results of a detailed simulation program CORSIKA, that the superposition assumption is not fulfilled for higher moments of the distributions of the main characteristics of extensive air showers. We will discussed some  extensions of the simply superposition model that ensure consistency with the results of detailed simulations.

\end{abstract}

\maketitle
\section{Introduction \label{intro}}

The primary cosmic radiation consists mainly of hydrogen nuclei (protons), but alongside them, as we have known since the middle of the last century, there are nuclei of heavier elements. Nuclei heavier than iron no longer make a significant contribution to the total flux, although they are important targets of studies on the origin and propagation of cosmic radiation. 

The steep cosmic ray energy spectrum causes that the flux of secondary particles reaching the earth's surface is determined by the flux of primary particles with relatively low energies. They interact with the nuclei of atoms of the Earth's atmosphere and when they exceed a certain, not very high threshold energy, they may initiate cascades of secondary particles. And if in the majority of cases these cascades fade away during the propagation through the atmosphere, if there is only a slight probability that something from them will reach the surface of the Earth, after having taken into account a significant gradient of the energy spectrum,  these highly fluctuated cases constitute the majority of incoherent particles, or particles associated in small groups - very small cosmic ray showers observed at the ground level.
Fluctuations in the development of Extensive Air Showers (EAS) depend on many factors. One of these is the mass of the primary particle. Understanding how it influences the development and fluctuations of EAS can greatly facilitate, and increase the accuracy of calculations as a tool for the interpretation of very small cosmic ray showers or single secondary cosmic ray particle flux measurements.  

The fundamental role in the shower development is played by inelastic interactions of primary cosmic ray particles. 
As a result, secondary hadrons are created, mainly pions, and among them about 1/3 are neutral pions, which immediately decay into two gamma quanta initiating electromagnetic cascades. Electrons (positrons) and photons make up the vast majority of particles in showers. Their interactions, which are described by quantum electrodynamics, are not a serious physical problem, although it would be a mistake to think that we know everything about them. The main problem of the simulations of the high energy primary cosmic cascades is the number of its constituents going into billions, which makes precise simulation calculations difficult and partly even impossible. Hadrons, nucleons in atomic nuclei interacting strongly do not pose such a problem. In this case, an important limitation is a lack of knowledge of physics of multiparticle production processes. Most often we are dealing with a small four-momentum transfer soft secondary particle creation.  It is described by more or less phenomenological models of particle production extrapolating available data from accelerator experiments by many orders of magnitude in energy and to areas of momentum immeasurable in collider experiments.

In the lowest energy region of the cosmic ray particle energy spectrum, which is the field of our main interest in this work, both the description of the electromagnetic component and the phenomenology of strong interactions, due to the multitude of data collected over the years, the situation apparently looks much better. Simulation description of physics of small showers (we will define hereafter ``small showers'' those induced by particles of energies below the ``knee'' $< 10^{15}$ eV) does not seem to cause any particular difficulties. However, there appear other specific problems connected with the very steep slope of the cosmic ray energy spectrum. As it is commonly known, the spectrum is power-law with index (differential) of about 3. This spectrum is imposed by Nature as an {\it a priori} factor in Bayesian approach whenever we want to compare the results of simulation calculations, which, out of obvious necessity, are made for a fixed energy of the primary particle, with measurements, which are just an {\it average} taking into consideration all {\it a priori} factors, and, therefore, also the steepness of the energy spectrum. 

This problem of course exists for all energies, but the influence of any fluctuations for higher energies is taken into account relatively easily. By virtue of the law of large numbers, we expect that the fluctuations will have a normal (possibly log-normal) distribution with a small relative width and an appropriate integration with a power spectrum gives a certain shift on the energy scale without significantly changing the nature of the observed relations. In the case of small and very small showers in extreme cases we deal with distributions of observables which are small natural numbers or distributions which are not normal and have additionally significant relative width (dispersion). Due to the steepness of the energy spectrum in cases of small showers, when integrating (averaging with the spectrum as a {\it prior}) we practically do not operate on average values, and sometimes we are quite far from them. To take such great fluctuations into account in simulation calculations, it is necessary to simulate a sufficiently great number of events and, although simulating a shower initiated by particles with energies of the order of TeV does not take much computing time, however, when such simulations must be repeated millions of times we stand in a situation similar to the calculations for ultra high-energy cosmic rays.

\section{CORSIKA program}
Modelling of strong interactions has a long history. Majority of models currently available on the market were developed a long time ago and are continuously refined and improved as new data become available. These models have to be implemented into the geometrical structure of EAS particles transport through the Earth's atmosphere. Complementing them with a much better understood formalism of electromagnetic cascades and descriptions of other more and less relevant processes leads to processes that simulate the development of EAS. One of the most commonly used programs for this purpose today is the CORSIKA  \cite{1998cmcc.book.....H,CORSIKA2020} developed over 30 years ago in Karlsruhe for the KASCADE experiment \cite{ANTONI2003490,doi:10.1142/97898145290440017}.  Since then this program has been significantly extended and developed. Over the years, new features have been added and the computational capabilities have been extended to meet the computational needs of new cosmic ray experiments and it is now used also for simulations at the highest observed energies (even up to 10$^{21}$ eV).

The CORSIKA program provides users with a number of options. They concern both the structure of the simulation algorithm itself, including models of high- and low-energy interactions, which should be attached to the basic simulation scheme, as well as those connected with the parameters of the simulated showers, which are of interest to the user (Cherenkov radiation, radio emission, and atmospheric neutrinos). There are so many options that just listing them with the minimum necessary description takes almost 200 pages in the latest version of the CORSIKA manual 
\cite{CORSIKA2020}.
The already linked program requires run parameters to be set using the control cards. In principle, for a typical simulations the default set embedded in the program itself is selected. However, if you want to use the CORSIKA program for a purpose slightly or more different than typical, you have to set the parameter values particularly carefully.
For the purposes of this work, we mainly want to use the program to simulate small and very small showers. For example, in our case, it is essential to determine precisely the size of the fluctuations we are dealing with at the lowest energies, so we must not use any thinning option. 

\section {Superposition model}

Most of the observed characteristics of EAS depend not only on the energy of the primary particle, but also on its mass (mass number). The primary cosmic ray flux is dominated by the nuclei of hydrogen and helium, the CNO group, a widely blurred group of 'intermediate' nuclei: Ne, Mg, and Si, and the iron group, never the less we know that the whole Mendeleev table can come from above. The mass composition of the primary cosmic radiation is known (in the small shower energy region) with a good accuracy, but anyway the mass of the primary particle is another factor to be taken into account when averaging simulation calculations, which increases their calculation time several times. 

In order to get rid of at least this problem, for almost half a century in the EAS physics a quite obvious model called the {\it superposition model} has been functioning \cite{Beer_1966,Beer_1968}. Since the atomic nucleus from the point of view of high-energy processes of particle production is a collection of nucleons (protons and neutrons, which do not differ at all in the aspect under discussion) it can be treated this way and one can assume that the EAS initiated by the nucleus of, for example, iron is a sum, a superposition, of 56 proton showers, where each proton would have 56 times less energy than the nucleus of iron as a whole.

This undoubtedly attractive model made it possible to formulate at once, as if by definition, many theorems concerning the comparison of, e.g., iron and proton initiated showers, which have been used and are still in use for the analysis of the mass composition of the primary cosmic rays in experimental studies.

\section {Average EAS characteristics}

\subsection{Shower size}

First of all, from the very definition there is a relation between the number of particles in proton and iron showers:

\begin{equation}
    N_{e,\mu} ({\rm iron}, E)=  56\times N_{e,\mu} ({\rm proton}, E/56) \label{sred}
\end{equation}

\begin{figure}
\centerline{
\includegraphics[width=0.5\textwidth]{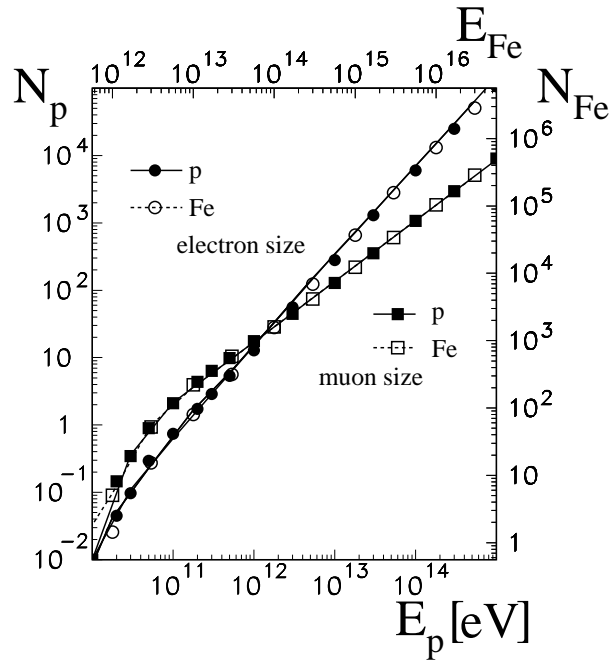}}
\caption{Average values of the number of electrons (circles) and muons (squares) at the sea level in vertical shower obtained using CORSIKA as a function of energy per nucleon of the primary particles the lower scale (the total proton energy of the proton showers). The upper energy scale shows the total energy of the iron nuclei. The graph shows at the same time results for proton showers (filled symbols and the left axis for the number of particles) and scaled-up iron showers (open symbols and the right axis). The solid lines show, respectively, the results of CORSIKA fits to the proton shower sizes used to obtain the results predicted by the superposition model for iron showers.
\label{sre}}
\end{figure}

\noindent
(surely, an analogous relation holds for showers initiated by any nuclei with other masses). $N_{e}$ and $N_{\mu}$ denote here the number of electrons (and positrons) and muons at the observation level, which will be hereafter called electron and muon shower size.
This relationship is illustrated in Fig.\ref{sre}.

As we can see, an obvious agreement occurs not only in the area of relatively high energies, where we are dealing with purely power-law relations, but also in areas where the average numbers of shower particles are smaller than 1, which means that there is a significant number of showers that do not reach the observation level (in our case the sea level) at all.

\subsection{Shower particle radial distributions}

It has been known for many years that the transverse distributions of particles in extensive air showers are well described by a simple formula proposed by Greisen \cite{greisen}. 
Its validity was confirmed by theoretical considerations and numerical calculations with respect to electromagnetic cascades by Kamata and Nishimura \cite{10.1143/PTPS.6.93}, hence its commonly accepted name: Nishimura-Kamata-Greisen (NKG) function (which we adopted also for the muon distribution):

\begin{equation}
\rho_{e/\mu}(r)= {N_{e/\mu} \over 2 \pi r_0^2}\: {\Gamma(4.5-s) \over \Gamma(s)\:\Gamma(4.5-2s)}\left({r\over r_0}\right)^{s-2}
\left({1+{r\over r_0}}\right)^{s-4.5}~~~,
\label{nkg}\end{equation}

\noindent
where $N_{e/\mu}$ is the electron/muon shower size, $r_0$  is a radial scale parameter 
called in the case of the electromagnetic cascade theory the Moli\`ere unit equal to about 2 radiation lengths units above the observation level (for muon distribution its value was adjusted and parametrized using the CORSIKA results) and $s$ is the age parameter. 

\begin{figure}
\centerline{
\includegraphics[width=0.33\textwidth]{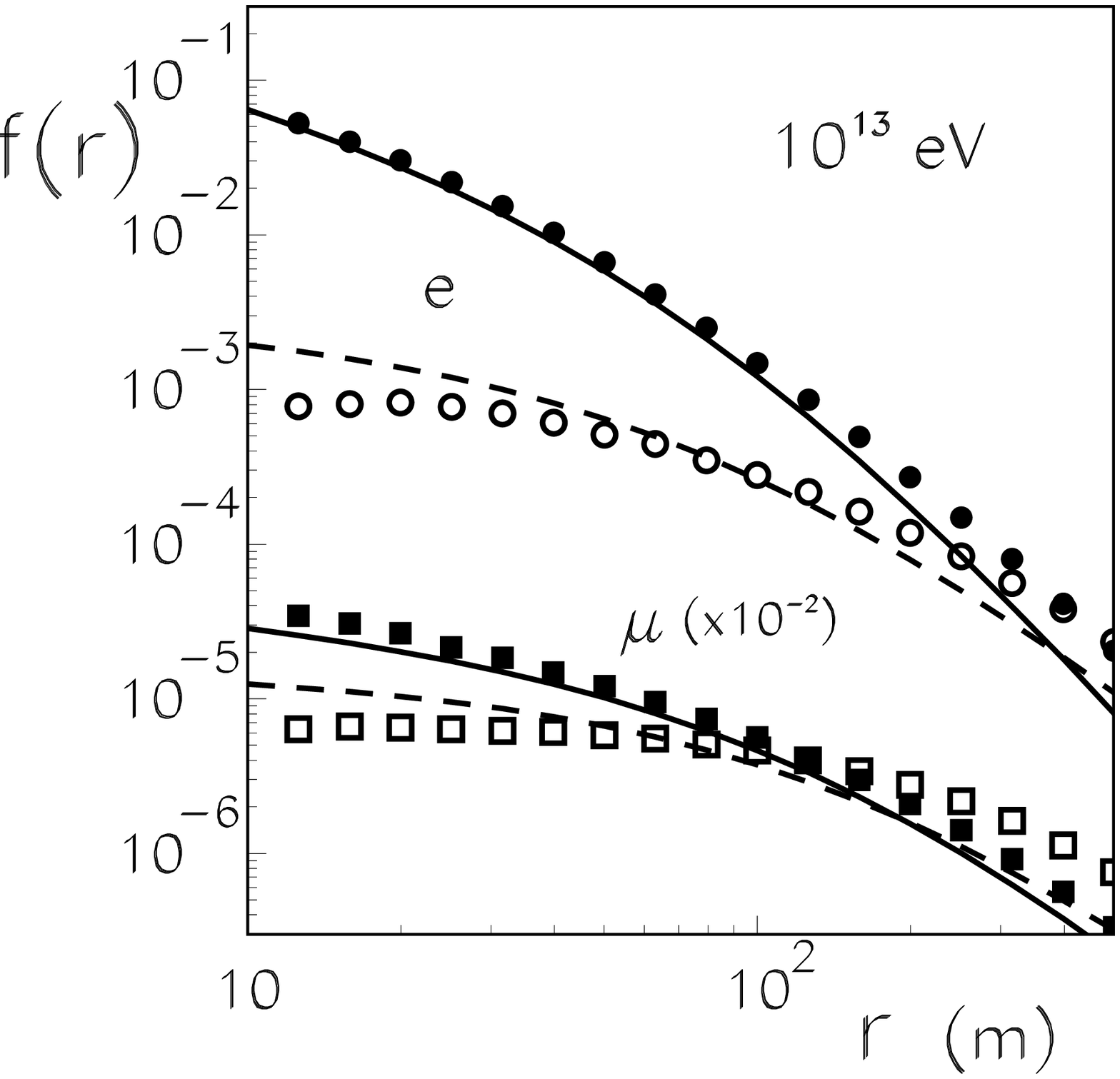}
\includegraphics[width=0.33\textwidth]{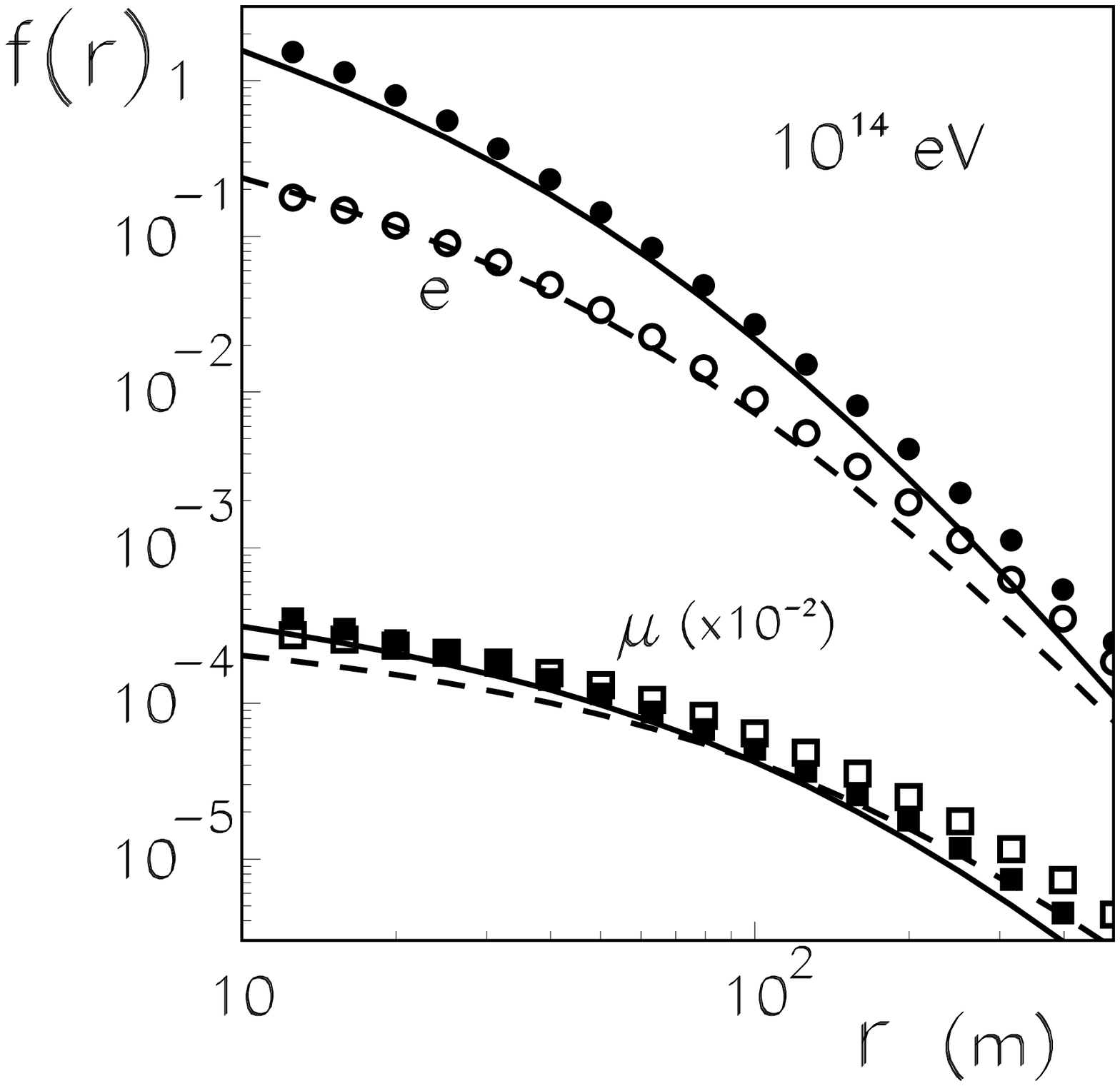}
\includegraphics[width=0.33\textwidth]{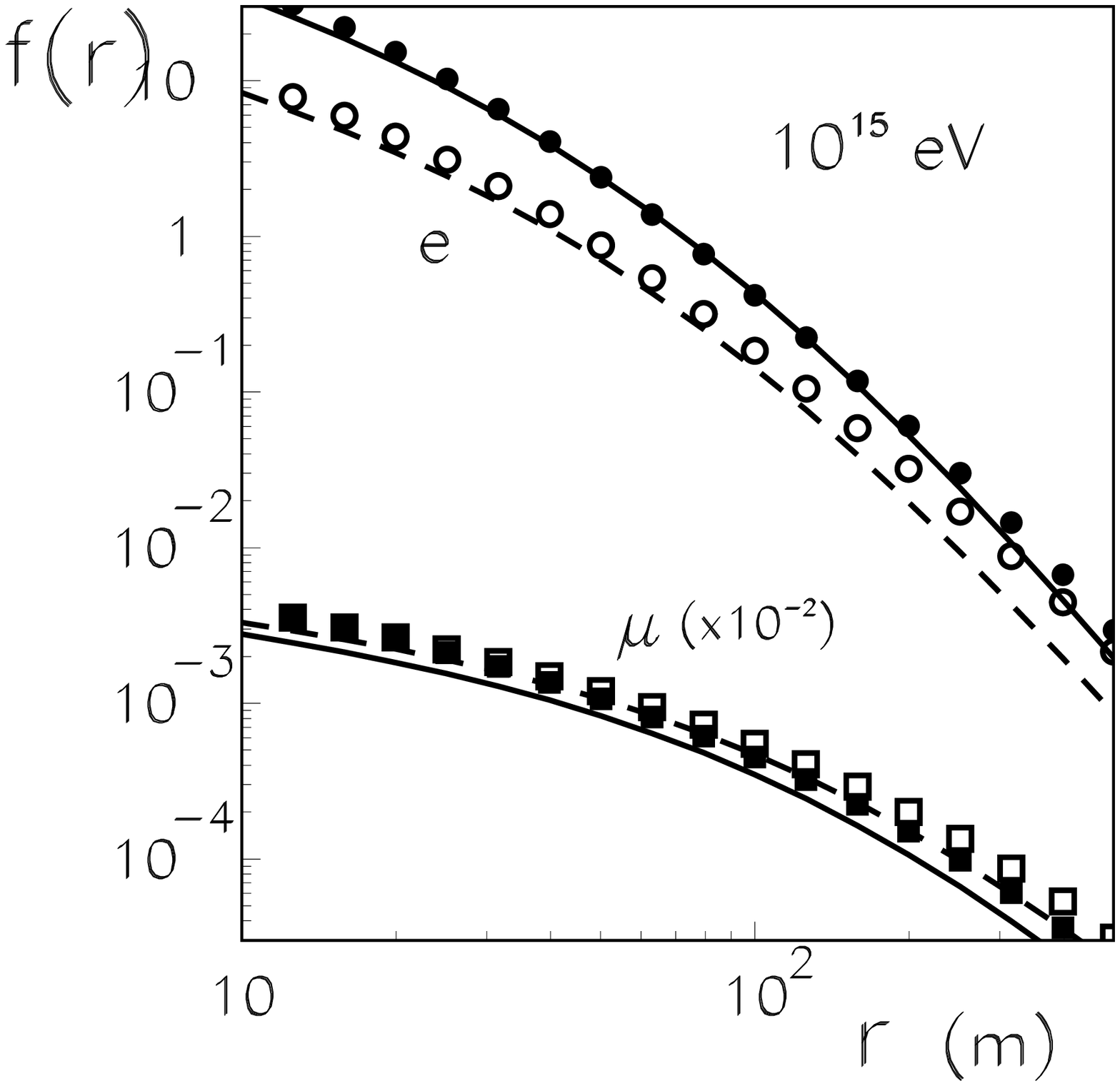}}
\caption{{Average radial distribution of electrons (circles) and mouns (squares) in CORSIKA showers of energy $10^{13}$ eV (left plot), $10^{14}$ eV (middle plot), and $10^{15}$ eV
(right plot) initiated by proton (black filled symbols) and iron nuclus (open symbols). Solid lines for proton induced showers, dashed for iron showers.}
\label{rad}}
\end{figure}

 Examples of average radial distributions for the electron and muon components of CORSIKA showers are shown in Fig.\ref{rad} for primary protons and iron nuclei of energies 10 TeV, 100 TeV and 1 PeV. The solid lines show the results obtained from a smooth parametrization of the age parameter $s$ in Eq.(\ref{nkg}) using the average electron and muon distributions for proton showers over the whole analysed energy range, also for very small showers. 
 
The results for iron showers shown in Fig.\ref{rad} with dashed lines have been obtained according to the superposition principle from the fits to correspondingly lower energy proton showers. The normalization is taken from the results shown in Fig.\ref{sre}.

For showers containing on average about 1 particle, which corresponds to a (mean) central density of 10$^{-6}$ per m$^2$ (100 GeV proton, -- Fig.\ref{sre}) the NKG function does not yet do a poor job of describing the mean distributions of electrons and muons, but at central densities of 10$^{-8}$ per m$^2$, which is the case, as seen in Fig.\ref{sre} for 10 GeV protons and iron nuclei with a total energy of about 1 TeV the distributions can be considered almost uniform and shower particles can appear almost everywhere up to a distance of several hundred meters from the shower axis.

For the purposes of this work, the lack of a statistically justified mean distribution of particles in very small showers is not of particular importance, since the integrals of these distributions we eventually aim for will be normalized by the total number of particles in such showers, which will be of the order of 0.01, as shown in Fig.\ref{sre}.

\subsection{Shower longitudinal profile}
The average number of particles in the showers and their average lateral distributions show in the low and very low energy region agreement with the superposition principle.

Analyzing the application of the superposition model in the case of small showers, it is worth to pay some attention to the longitudinal development of the showers.
The features of interest here in general are very difficult for any precise observations and measurements. Nevertheless, the comparison of the longitudinal development of showers resulting from the CORSIKA program in the case of proton and iron induced showers may provide important information regarding the considered superposition hypothesis.
The longitudinal profile of the shower provides an opportunity to follow the process of successive nucleon interactions of the initial heavy nucleus and the superposition mechanism. 

In the case of a shower initiated by a proton, its size at the maximum (the largest number of particles that the developing shower reaches at a certain point) is determined to a good approximation by the energy of the primary particle.

\begin{equation}
    N_{\rm max} = \frac 1 2 \frac{\langle n \rangle}{3}\: \langle K_{\rm inel}\rangle \:\frac{E} {\epsilon_c}
\end{equation}
where $\epsilon_c$ is the critical energy of electron in air, $\langle K_{\rm inel}\rangle$ effective interaction inelasticity coefficient and $\langle n \rangle$ is the effective pion multiplicity \cite{deile201113th}. 
It is a normalising factor for the longitudinal shower development function describing the number of particles at the depth $x$ in the atmosphere and can be described by different phenomenological formulas \cite{Matthews_2010}. We have decided to use the relation called the Gaisser-Hillas function \cite{1977ICRC....8..353G} in our superposition hypothesis examination. 

\begin{equation}
    N(x)= N_{\rm max} {\left( \frac{x-x_0}{x_{\rm max} - x_0} \right)}^{\left({ \frac{x_{\rm max} - x_0}{\lambda}} \right)}\:
    \exp{\left( - \:\frac{x - x_{\rm max}}{\lambda}\right)}\label{GH}
\end{equation}

In the case of heavy nucleus we are dealing with $A$ nucleons each of which can act at a specific depth in the atmosphere $x_i$ ($i=1, 2, 3, \ldots, A$).
If $F_A(\{x_i\})$ describes the $A$-dimensional probability density of the set of positions $x_i$, for the superposition model, when all $x_i$s are independent of each other, we have
\begin{equation}
    F_A\left(\{x_i\}\right)= \prod \limits_{j=1}^A  \frac{1}{\lambda_p} \:\exp\:(-x_j/\lambda_p) \label{fsh}
\end{equation}
where $\lambda_p$ is the interaction length of proton. 

It may be noted, obviously, that the averaged longitudinal profile of a shower initiated by a cosmic ray nucleus of mass $A$ is in this case equal to the sum of $A$ average profiles of proton showers of correspondingly smaller ($E/A$) energy only if  
interaction properties ($\langle K_{\rm inel}\rangle$, $\langle n \rangle$) and cross section ($\lambda_p$) does not change with the particle energy. 
 
This leads to the fact that the average depth of the maximum of the iron shower is exactly in the exact place of the maximum of the shower for the proton shower with the initial energy 56 times smaller. 
\begin{equation}
    x_{\rm max} ({\rm iron}, E)=   x_{\rm max} ({\rm proton}, E/56) \label{xm}
\end{equation}

\begin{figure}
\centerline{
\includegraphics[width=0.5\textwidth]{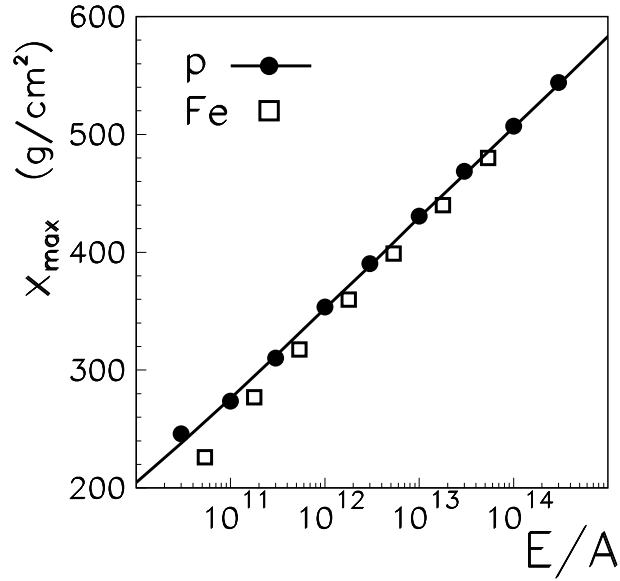}}
\caption{Average depth of the shower maximum obtained with the CORSIKA for primary proton (black circles) and iron (open squares) as a function of the primary particle energy per nucleon. The line is the fit for vertical proton showers.
\label{xmax}}
\end{figure}

The situation for the longitudinal development of CORSIKA showers is illustrated Fig.\ref{xmax}. As can be seen, a general agreement with Eq.(\ref{xm}) is maintained. A small deviation visible for very small showers is a result of the strong energy dependence of the interaction characteristics of nucleons with energies below a few TeV. This situation can be handled by the so-called elongation-rate theorem in its full form \cite{1977ICRC...12...89L,1981ICRC....6..234M}.

\section{EAS fluctuations -- higher moments}

\subsection{Shower size fluctuations}

\begin{figure}
\centerline{
\includegraphics[width=0.25\textwidth]{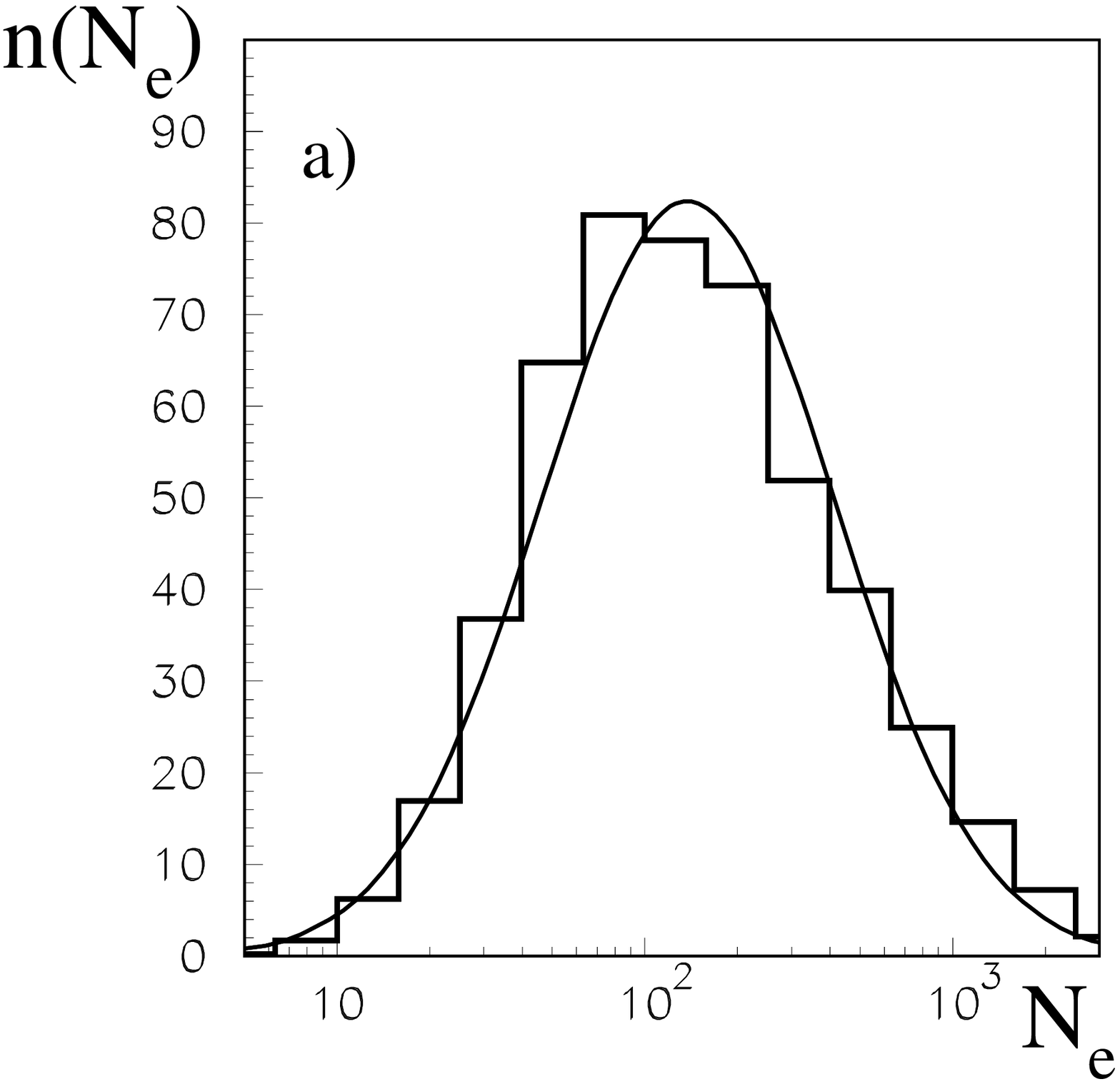}
\includegraphics[width=0.25\textwidth]{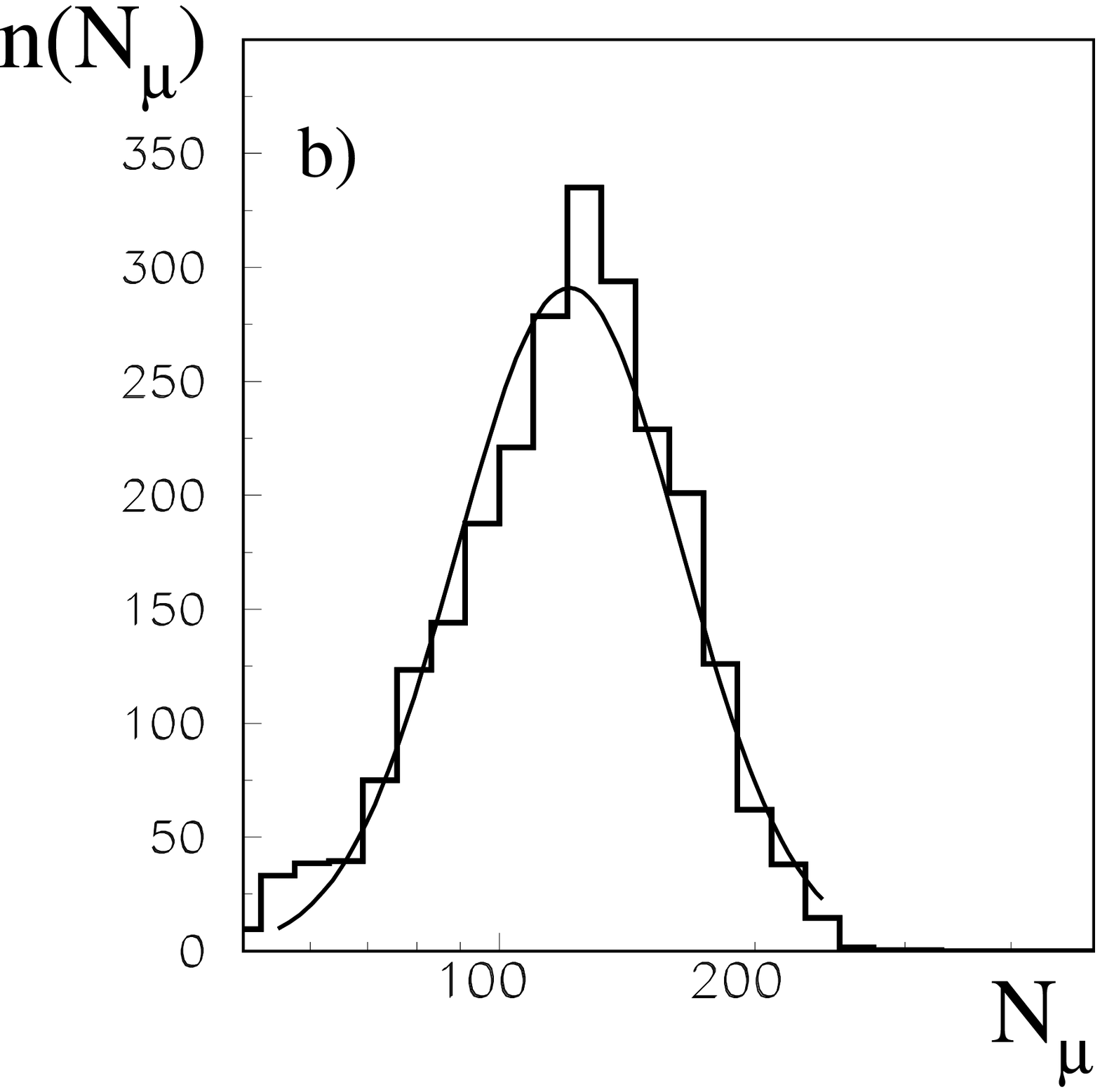}
\includegraphics[width=0.25\textwidth]{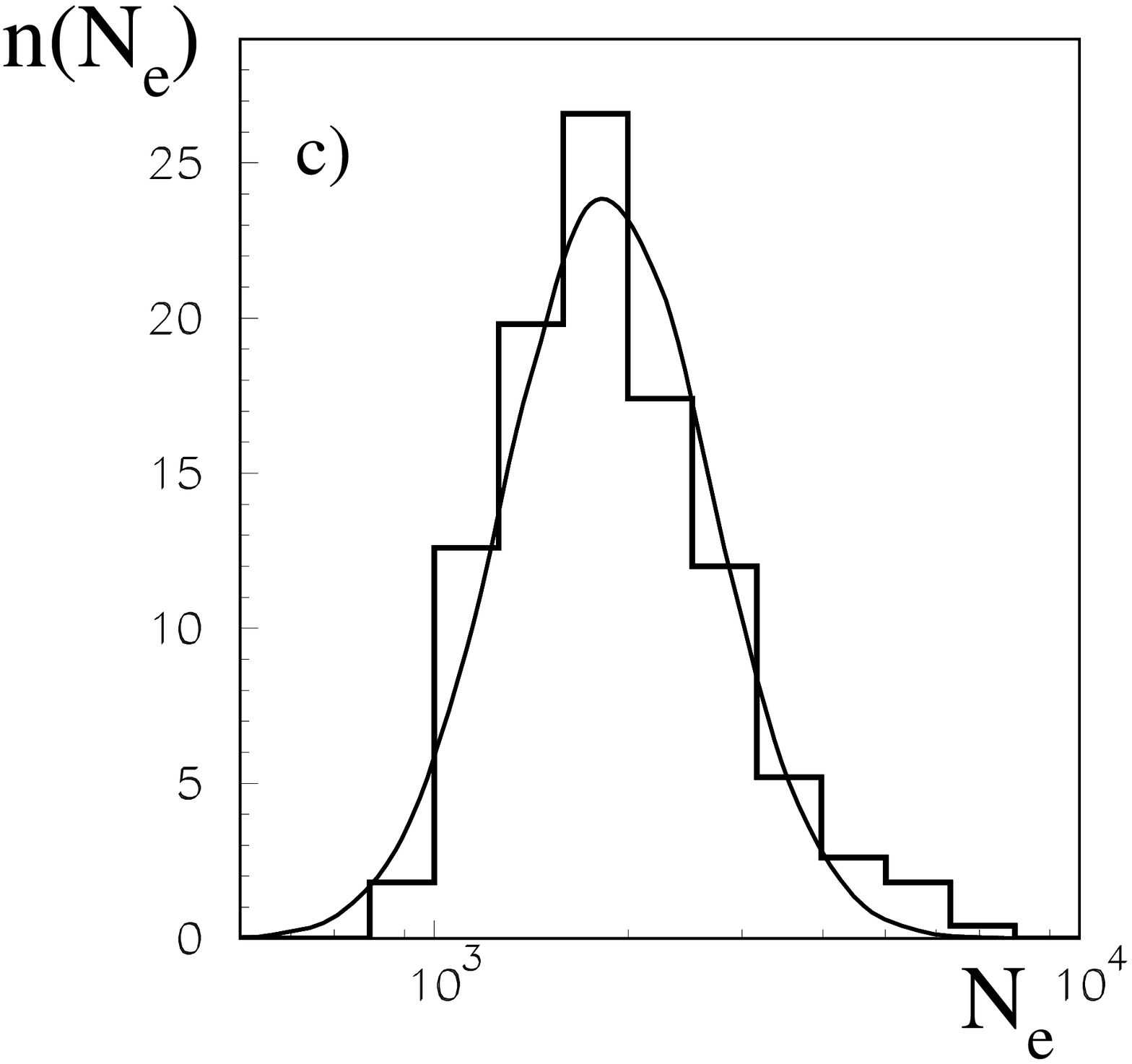}
\includegraphics[width=0.25\textwidth]{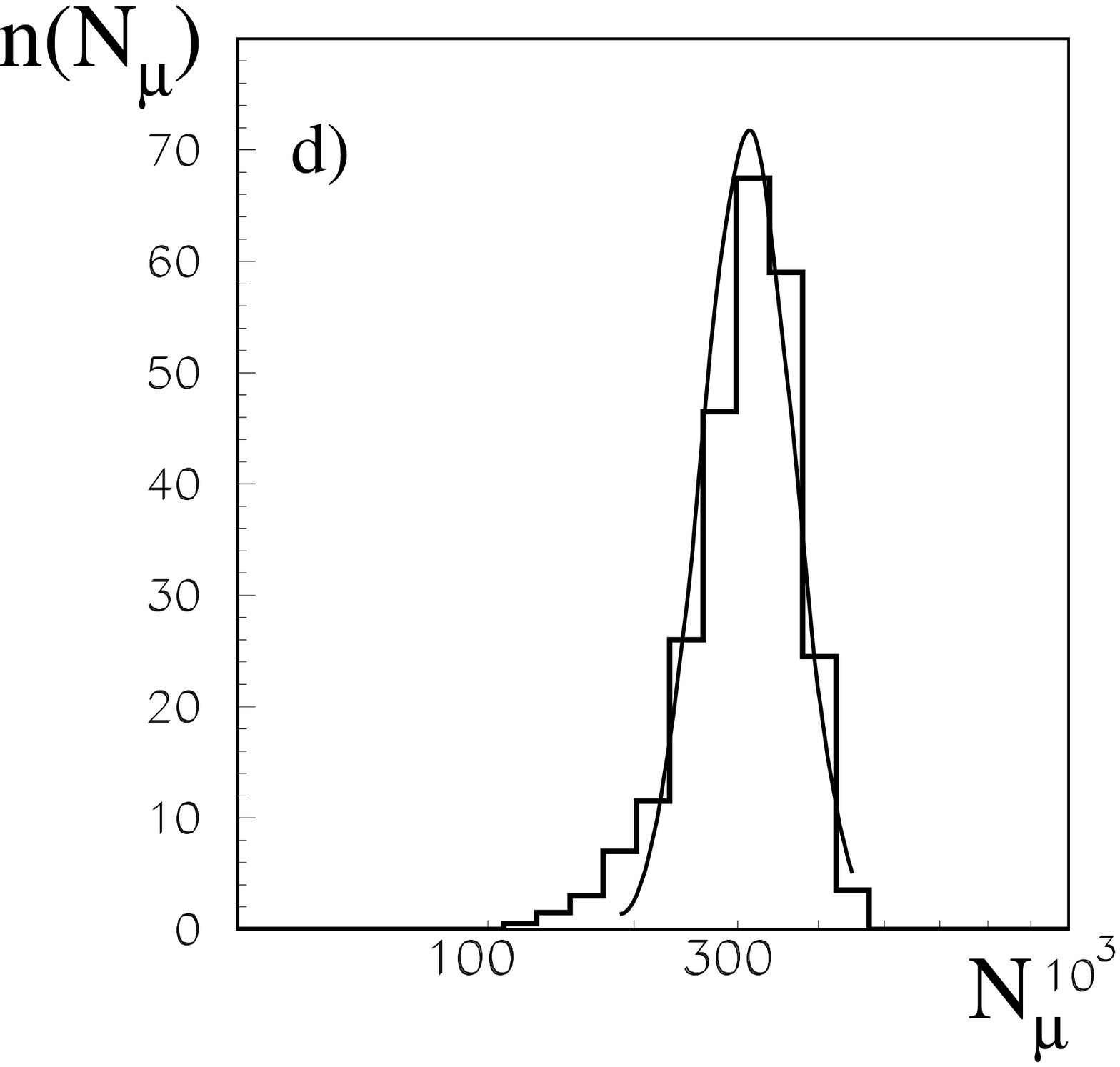}}
\caption{Examples showing spread of electron (a, c) and muon (b, d) numbers in CORSIKA vertical showers initiated by protons of energy of 10$^{13}$ eV (a, b) and 10$^{15}$ eV (c, d). Lines shows respective fits of the Gaussian (Log-normal) distribution.
\label{fluct}}
\end{figure}

For a fixed energy of the primary particle, the electron and muon size of EAS is surely not the same every time and we expect a fluctuation. These fluctuations are an intrinsic property of reality and of the simulations and result from the probabilistic nature of multi-particle production processes.  Fig.\ref{fluct} shows examples of electron and muon CORSIKA shower size spreads for showers initiated by protons with energies of 10$^{13}$ eV and 10$^{15}$ eV. The lines correspond to the Gaussian distribution (actually Log-normal) fitted to the histograms shown. As can be seen, in all cases the Log-normal distribution describes the scatter of sizes quite well. 

For showers smaller than those illustrated in Fig.\ref{fluct}, when the sizes are of the order of a few particles, the Log-normal distribution is obviously no longer the best. For such a small number of particles, we should expect an important correlation between them caused by their common origin from subcascades, which developed accidentally just above the level of observation. In addition, of course, in such cases, Poisson-type statistical fluctuations are superimposed on fluctuations in their number caused by the probabilistic nature of the phenomenon itself. 

For the superposition model, we have, taking into account that 56 proton showers consisting of one iron shower are independent from each other and each of them fluctuates as in the examples shown in Fig.\ref{fluct}, dispersions of the distribution of the total number of electrons (and muons) for the proton and iron EAS should satisfy the relation
\begin{equation}
    D_{e,\mu} ({\rm iron}, E)~=~  \frac{1}{ \sqrt{56}} ~~ D_{e,\mu} ({\rm proton}, E/56) \label{dys}
\end{equation}

\begin{figure}[bh]
\centerline{
\includegraphics[width=0.5\textwidth]{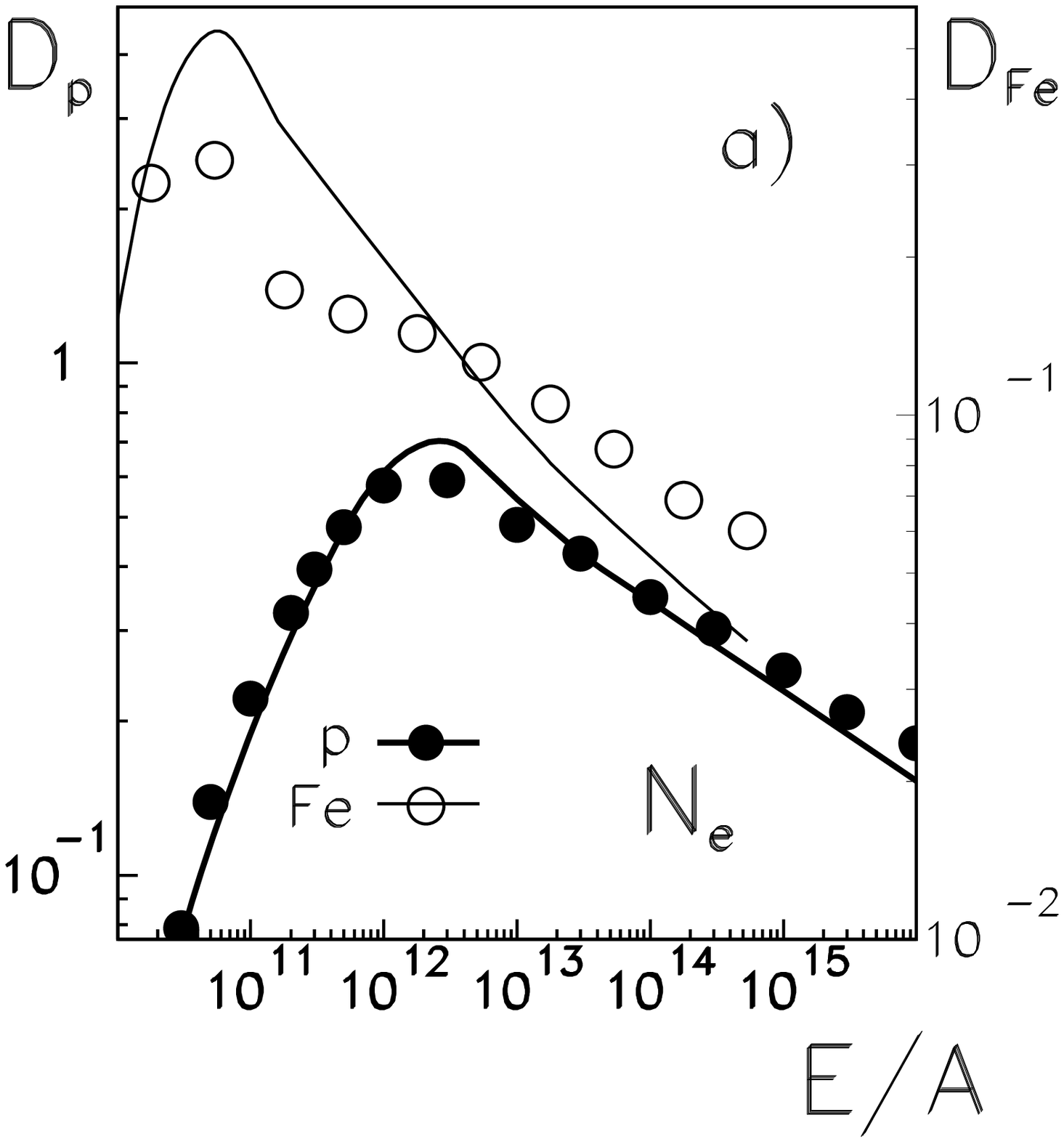}
\includegraphics[width=0.5\textwidth]{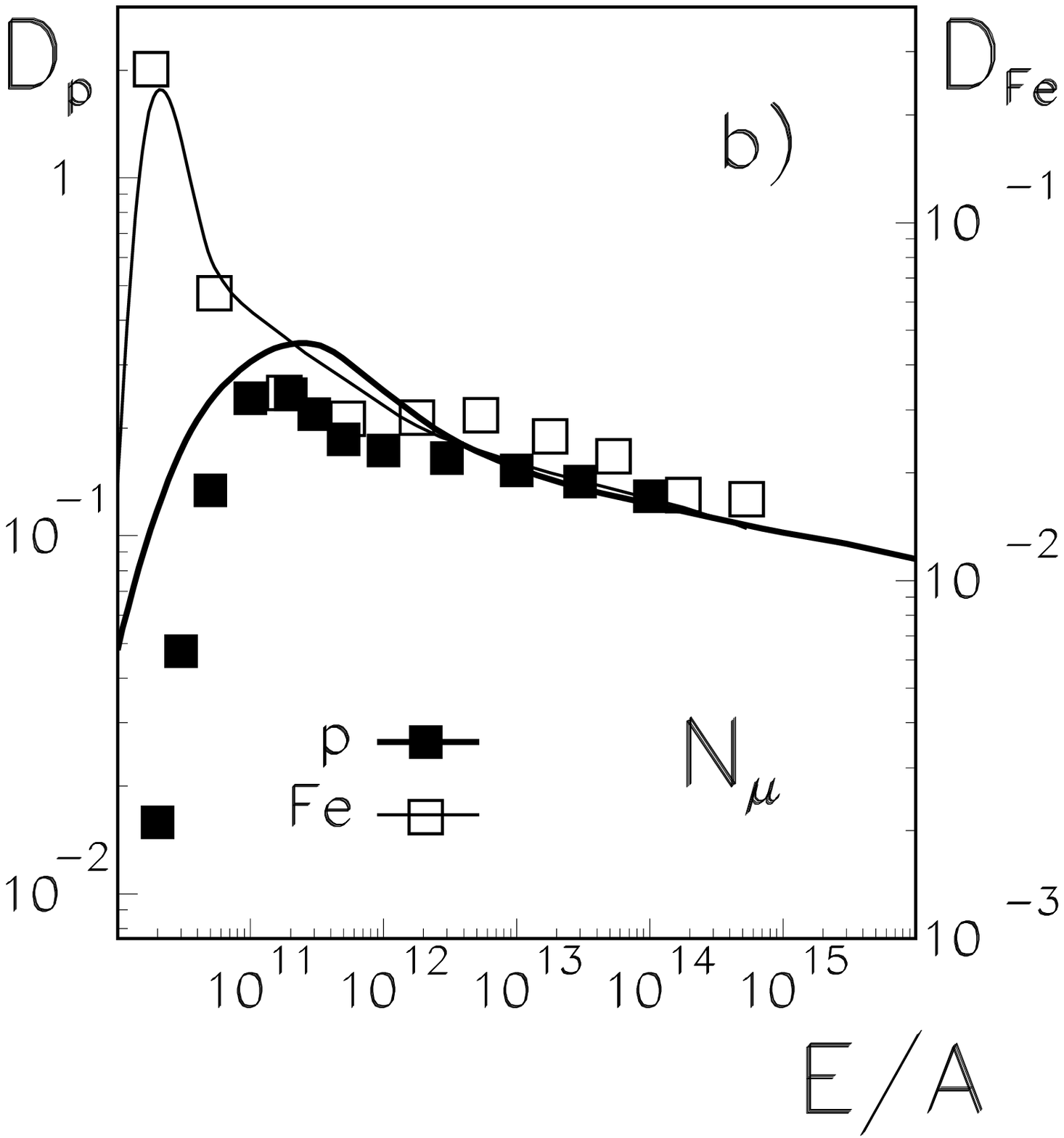}}
\caption{Dispersion of the number of electrons a) and muons b) in vertical CORSIKA shower as a function of the energy per nucleon of the primary particles. There are simultaneously shown results for dispersion of proton showers (black dots, and left axis) and scaled-up iron showers (open symbols, and right axis). The thick solid lines represent smooth fit for CORSIKA proton showers. Thin lines represent iron shower results of the superposition model with additional small statistics conditions taken into account.
\label{dyspe}}
\end{figure}

Fig.~\ref{dyspe} shows the CORSIKA proton shower results represented by black symbols compared with iron shower results (empty symbols). 
The first thing that is noticeable is that the scaling according to Eq.(\ref{dys}) does not work for CORSIKA showers. Especially in the case of electron size, Fig.\ref{dyspe}(a), the points corresponding to iron showers after rescaling corresponding to the relation Eq.(\ref{dys}) lie well above the points for proton showers.
The superposition model regarding the CORSIKA simulations shows clearly that fluctuations of the total number of electrons in showers initiated by iron nuclei are wider than one could expect (almost about a factor of 2 \cite{PhysRevD.25.2341,Schatz_1994}). In the case of the muon component this difference is also present, but it is less significant.

In Fig.\ref{dyspe} the thick lines show smooth fits to the CORSIKA results of dispersion in proton showers, while the thin lines show the results obtained from a superposition model that relates dispersion in iron showers to dispersion in proton showers (with additional small statistics requirements taken into account).
The superposition model (taking into account the necessary modifications for showers with very small number of particles at the observational level) tends, in general to the relation given in Eq.(\ref{dys}). Rescaled lines for iron showers are approaching lines for protons for muons already above the energies of iron nuclei of about a few times $10^{12}$ eV, and for electrons the agreement appears higher. However, the superposition model results clearly do not agree with the points CORSIKA simulations.

\subsection{Shower longitudinal profile fluctuations}

\begin{figure}
\centerline{
\includegraphics[width=0.5\textwidth]{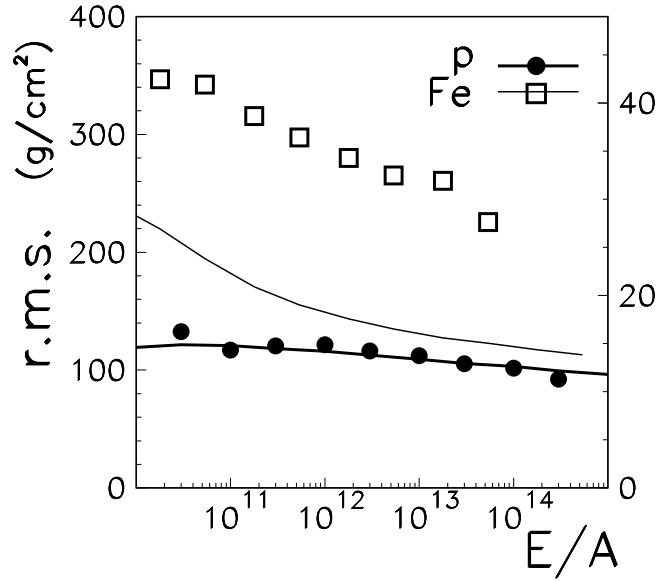}}
\caption{Dispersion of the depth of the maximum shower obtained with CORSIKA for primary proton (filled circles and left vertical scale for dispersion) and primary iron (empty squares and right-hand scale for the respective dispersion) as a function of primary particle energy per nucleon. The thick solid line is our fit to the proton shower results. The thin line represents result of superposition model of the iron shower with additional small statistics conditions taken into account.
\label{dysf}}
\end{figure}

A similar, but even more significant inconsistency is observed in the dispersion of the position of the shower maximum. The situation is shown in Fig.\ref{dysf}.
The CORSIKA results for proton showers are shown by black circles and at the same time the $\sqrt{56}$-scaled dispersions for showers initiated by iron nuclei are shown by empty squares. The superposition model Eq.(\ref{dys}) suggests that all these points for proton and iron showers should lie on the same line, but in CORSIKA reality they clearly do not.

The dispersion of proton showers $x_{\rm max}$ points are joined by a thick solid line. It is our fit to the CORSIKA points.
It is used to obtain results
for iron showers in the superposition model described mainly by the equation Eq.(\ref{fsh}) completed by the Gaisser-Hillas function Eq.(\ref{GH}).
According to the superposition model, the $\sqrt{56}$-scaled dispersion for showers initiated by iron nuclei should agree with the results for proton showers, and this is indeed the case. Superposition model results for iron showers shown in Fig.\ref{dysf} by the thin line clearly asymptotically tend towards the agreement with the results for proton showers. 

However, the problem is that CORSIKA shows a significant inadequacy of the simple model in this case: the thin line is far from connecting the empty squares in Fig.{dysf}.

The dispersion of $x_{\rm max}$ values obtained from the CORSIKA program for iron showers are about three times larger than expected for the superposition model (at the same energy per nucleon). This is a clear signal that the simple superposition model needs to be modified.

\section {Modifications of the simple superposition model}

The simple superposition model in the case of average $N_e$ and $N_\mu$ must lead to the relation given by Eq(\ref{sre}) and for $x_{\rm max}$ by Eq.(\ref{xm}) (neglecting the mentioned energy dependence of interaction parameters of particles) with the  regardless of whether the nucleon sub-showers comprising the iron shower are in any way dependent on each other or not. 
From the physical picture of the hadron cascade development process, it is clear that the subshowers are correlated to some extent. The interaction cross section of a heavy nucleus with the nucleus of the atmosphere is clearly larger than the cross section of a single nucleon interaction. The widely accepted Glauber model \cite{10.1007/978-1-4684-1827-9_43,GLAUBER1970135} of nuclear collision and its further evolution to the form of the ``wounded nucleon'' model \cite{Bialas:1976ed} explain the difference in cross sections by the fact that a single interaction of nuclei involves several individual nucleon-nucleon interaction at a time. The average number of interacting, wounded nucleons is given by the ratio of active cross sections:

\begin{equation}
    \langle w_{A}\rangle= A\:
    { \sigma_{p-{\rm air}} \over \sigma_{A-{\rm air}}} 
    = A\:{\lambda_{A-{\rm air}} \over \lambda_{p-{\rm air}}}
    \label{wou}
\end{equation}

\begin{figure}
\centerline{
\includegraphics[width=0.5\textwidth]{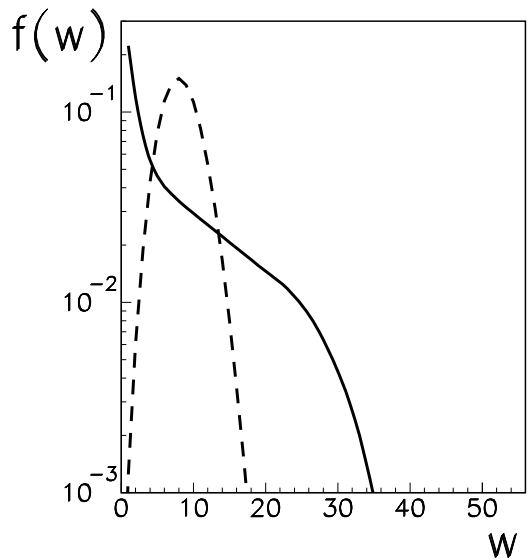}}
\caption{{Distribution of number of wounded nucleon for Fe-N interactions at the energy of 1 TeV/nucleon calculated for microscopic Glauber picture \cite{PhysRevD.46.5013,Wibig_1998} (solid line) and the respective binomial distribution with the same average number of wounded nucleons shown by the dashed line.}
\label{w}}
\end{figure}

The number of wounded nucleons fluctuates
and its distribution is in principle unknown.
As the first approximation we used the assumption of independence of interactions of nucleons in the nucleus. Thus, the actual number of wound nucleons must follow a binomial distribution. We called this solution the { \it wounded nucleon superposition model}.

In the geometrical concept of the interaction of compound nucleons, which in some sense underlies the  Glauber theory, the individual collisions of nucleons from a beam nucleus with some nucleon of a target nucleus are not, and cannot be, independent. 
The distribution of the number of wounded nucleons can be attempted to be determined using a simple geometric picture, but the assumption that the number of wounded nucleons depends on the integrated transparency of the two colliding nuclei may not necessarily be correct and accurate. 
Detailed calculations \cite{PhysRevD.46.5013,Wibig_1998}
show that the actual width of the distribution of the number of wounded nucleons is much wider than would be implied by the model of independent interactions with the same mean defined by the transfer Eq.(\ref{wou}). This is shown in Fig.\ref{w}.

Taking into account these more realistic distributions we obtain an increase in the spread of the some shower characteristics. This applies both to the total shower sizes (muon and electron) and to the width of the distribution of the depth of the shower development maximum $x_{\rm max}$. We called this version of the superposition  model {\it genuine wounded nucleon superposition model}.

Following the wounded nucleon idea, we need to modify Eq.(\ref{fsh}) to take into account the microscopic mechanism of nuclei interaction.  If in the particular interaction the number of wounded nucleons is equal to $w$,  then just $w$ nucleons will interact in the same place, at the same depth in the atmosphere, leading to the multiplication of identical (or almost identical) subshowers. 

We must also assume something about the interaction of the remaining not-wounded nucleons. Let us assume, as a first considered model, that the nucleus in the first interaction fragments itself completely and all nucleons of which it was composed travel further in the atmosphere as free protons. 
Then

\begin{equation}
    F_A\left(\{\overbrace{x_0, x_0,...,x_0}^{w\: times},x_{w+1}, x_{w+2}..., x_A\}\right) = \frac{1}{\lambda_{Fe}^w} \:\exp\:\left(-\:\frac{ w \: x_0}{\lambda_{Fe}}\right) 
    \prod \limits_{j=1}^{A-w}  \frac{1}{\lambda_p} \:\left.{\exp\left(-\:\frac{(x_j-x_0)}{\lambda_p}\right)}\right|_{x_j>x_0} 
    \label{fshdwa}
\end{equation}
\noindent
and if we also consider Eq.(\ref{wou}) we immediately obtain, that the distribution of the value of the depth of the first interaction of the single nucleon $x_i$ (we neglect the unimportant ordering occurring in Eq.(\ref{fshdwa}))  is the same as in the model of the simple superposition described by Eq.(\ref{fsh}) \cite{PhysRevD.46.5013}. 
It follows that the position of the EAS maximum, as determined, first of all, by the depths of the first interactions, will be the same in the simple superposition model and in the wounded nucleon model assuming the first interaction of the nucleus as common for $w$ its nucleons and independent cascades of remaining $A-w$ non-wounded  nucleons starting at  the first nucleus interaction point.
The correlation introduced in Eq.(\ref{fshdwa}) also, of course, does not affect the average shower size values and the averaged electron and muon radial distributions.

\begin{figure}
\centerline{
\includegraphics[width=0.5\textwidth]{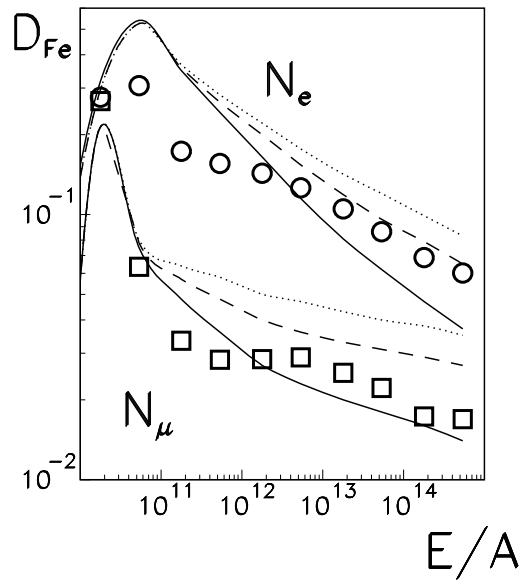}}
\caption{
Dispersion of the electron and muon size of CORSIKA iron showers (shown already in Fig.\ref{dyspe}) as a function of energy per nucleon. 
Thin solid line represents results of the superposition model (as in Fig.\ref{dyspe}), dashed line wounded nucleon superposition model, and dotted line genuine wounded nucleon superposition model, respectively.
\label{mod1}}
\end{figure}

The correlation introduced by the wounded nucleon superposition model, which binds independent, in the simple superposition model, sub-cascades of single nucleons, however, must lead to an increase in the fluctuations of both electron and muon average shower sizes and to an increase in the spread of the position of the shower maximum in showers initiated by heavier nuclei. This effect is shown in 
Fig.\ref{mod1} where we show the dispersion of electron and muon sizes for iron induced showers. The results for the simple superposition are shown by the solid line (same as in Fig.\ref{fluct}) and for the wounded nucleon superposition model by the dashed line.

\begin{figure}
\centerline{
\includegraphics[width=0.5\textwidth]{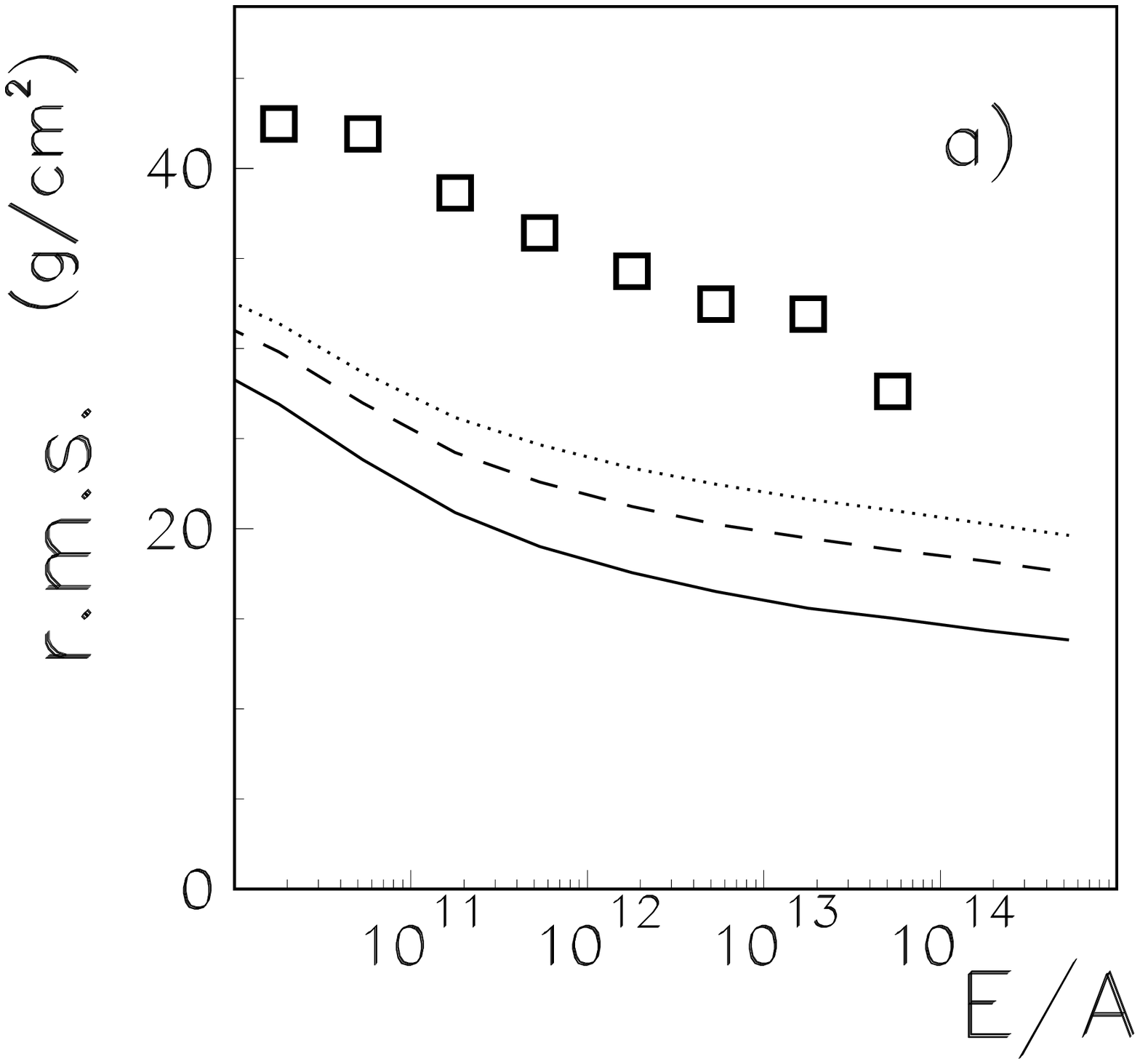}
\includegraphics[width=0.5\textwidth]{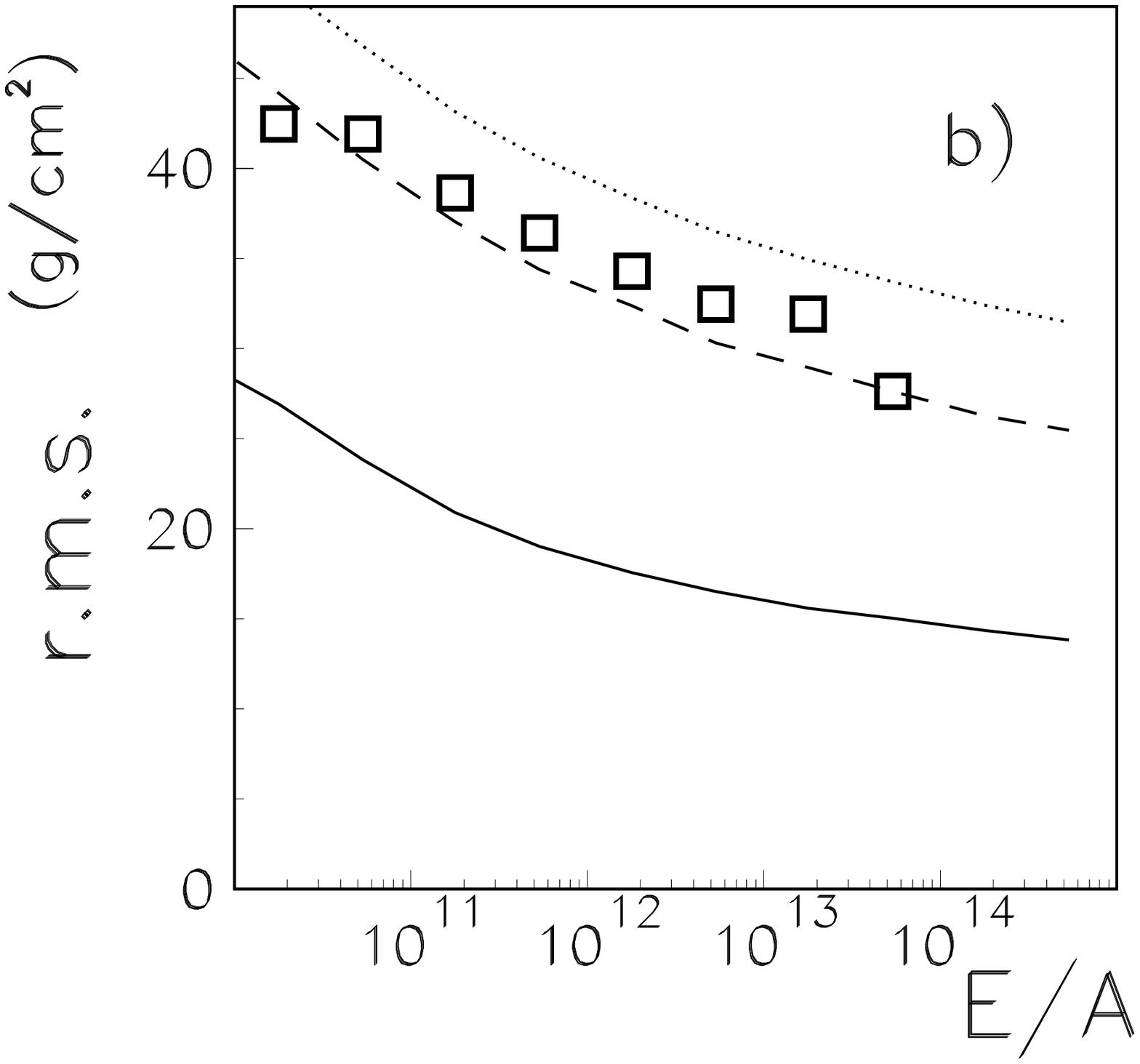}}
\caption{{
Dispersion of the depth of the shower maximum obtained with the CORSIKA for primary iron as a function of the energy per nucleon for total fragmentation of the primary nucleus in the first interaction (a), and assuming no fragmentation of the nucleus consisting of non-wounded nucleons (b). Solid lines are obtained using simple superposition model, dashed represents the wounded nucleon superposition model and dotted lines are the result of the genuine wounded nucleon superposition model.}
\label{moddxmax}}
\end{figure}

The increase in dispersion for the position of the shower maximum resulting from the introduction of the correlation described above is about 5 g/cm$^2$ and is shown by the dashed line in Fig.\ref{moddxmax}a. It does not explain the effect observed for CORSIKA showers.

The assumption of a complete fragmentation of the nucleus in its first interaction obviously does not have to be correct. Another extreme case, when non-wounded nucleons remain as one smaller nucleus, does not change the situation given in  Eq.(\ref{fshdwa}) substantially, if we only consider the average values of the discussed shower parameters. Replacing the product in Eq.(\ref{fshdwa}) by a complex of appropriate exponential factors for successively smaller nuclei remaining after successive interactions leads also to the same distribution of the first interaction positions. The contribution to the development of EAS from successive interactions of fragmented or unfragmented remnants of primary nuclei has no significant effect either on the size of the shower or on the depth of its maximum.
However, some effects are expected for higher moments of the distributions of some characteristics of the shower parameters. 

The dispersion of the flux distributions for the model with no fragmentation is shown in Fig.\ref{mod1} by the dashed lines. The changes in the widths of the shower size distributions are not large. They reach 10-20\% as it can be seen in Fig.\ref{mod1}. However, the change is great when considering the characteristics of the longitudinal distribution of the iron showers. A comparison of Fig.\ref{moddxmax}a and Fig.\ref{moddxmax}b shows that the differences are as high as factor 2.

Of course, both assumptions about fragmentation of heavy nuclei are extremes and the truth most likely lies in the middle. Our calculations show that in the CORSIKA program it is closer to the situation of reduced fragmentation. 
The binomial distribution of the number of wounded nucleons even in the absolute lack of fragmentation does not lead to the seen in CORSIKA showers width of the distributions of $x_{\rm max}$. A more realistic distribution following the Glauber model predictions provides a sufficient margin for reduced fragmentation models.

\section{Conclusions}

We used the CORSIKA program to obtain shower characteristics from the lowest primary energies allowed by the program. The average values of the most important parameters for shower: the total number of particles, the shape of their lateral and longitudinal distributions and their second moments, were found and analysed. We did this separately for the soft component - electrons and for the hard component - muons. 

Analyzing the electron and muon shower sizes, radial and longitudinal shower particle distributions for different atomic masses of the primary nuclei, we have examined the superposition hypothesis, saying that the extensive air shower produced in the interaction of a nucleus of mass $A$ is a simple compound of $A$ showers initiated by nucleons (protons) of appropriately lower energy. 
The behaviour of the average shower size values agrees with this assumption. 
 However, investigating fluctuations of the shower size (electron and muon), we have noticed that the assumption that the shower from nucleons composing according to the superposition principle, independently of each other, does not correspond to reality, or at least does not correspond to the reality of the CORSIKA program. 
 
 We have investigated possible modifications to remedy this inconsistency by including a correlation between some of the constituent subshowers.
 
We have tested the possibility that the fluctuations in the number of wounded nucleons are independent, and therefore described by a binomial distribution and a more realistic case distribution that more closely approximates the collective geometric picture of the interaction. In both cases we obtained better agreement with the results for the CORSIKA showers, but the most sensitive shower parameter for resolving between these possibilities turned out to be the position of the shower maximum, or more precisely the dispersion of this quantity. 

The magnitude $x_{\rm max}$ for very small shower energies is unfortunately difficult to achieve, or not achievable experimentally at all. Apart from the characteristics of the interaction act itself and interaction cross sections, it depends also on the fragmentation model of heavy nuclei initiating the showers. We have tested two extreme possibilities: either the nucleus completely fragments already in the first interaction, or the non-wounded nucleons propagate further in the atmosphere as one smaller nucleus.The CORSIKA results suggest that the latter case is closer to reality while allowing for the possibility of limited fragmentation, which seems to be in accordance with expectations.

\section*{References}
\bibliography{superposition2}

\end{document}